\documentclass[pre,aps,floatfix,twocolumn,superscriptaddress,showpacs]{revtex4-1}
\usepackage{color}
\usepackage{graphicx}
\usepackage{longtable}
\usepackage{amsmath}
\usepackage{amsthm}
\usepackage{amssymb}
\usepackage{lipsum}
\usepackage{latexsym}
\usepackage{bm}

\newcommand{\mbf}{\mathbf}

\graphicspath{{./figures/}}

\begin{document}
\date{\today}
\title{Time correlation functions in the Lebwohl-Lasher model of liquid crystals}
\author{Anoop Varghese}
\email{a.varghese@reading.ac.uk}
\author{Patrick Ilg}
\email{p.ilg@reading.ac.uk}
\affiliation{School of Mathematical, Physical, and Computational Sciences, University of Reading, Whiteknights, Reading RG6 6AX, United Kingdom}
\begin{abstract}
Time correlation functions in the Lebwohl-Lasher model of nematic liquid crystals are studied using theory and 
molecular dynamics simulations. In particular, the autocorrelation functions of angular momentum and nematic director fluctuations are calculated in the long-wavelength limit.
The constitutive relations for the hydrodynamic currents are derived using a standard procedure based on non-negativity of the entropy production.
The continuity equations are then linearized and solved to calculate the correlation functions. We find that the transverse angular momentum fluctuations are coupled to the 
director fluctuations, and are both propagative. The propagative nature of the fluctuations suppress the anticipated hydrodynamic long-time
tails in the single-particle autocorrelation functions. The fluctuations in the isotropic phase are however diffusive, leading to $t^{-d/2}$ long-time tails in $d$ spatial dimensions.
The Frank elastic constant measured using the time-correlation functions are in good agreement with previously reported results.
\end{abstract}
\date{\today}
\maketitle
\section{Introduction}

Nematic liquid crystals are formed of elongated or disc-like molecules and are characterized by spontaneous long-range order in the orientation of the constituent molecules~\cite{prost_book,chandrasekhar_book}.
The ordering occurs either at low temperatures or at high density, across a first-order transition point from an isotropic phase.
The orientational order is also strongly responsive to external electric fields, resulting in field-dependent light transmittance -- a property which find applications in a spectrum of display devices.
From a fundamental point of view, being the intermediate state of matter between solids and liquids, liquid crystals have been of extensive research in the last several decades~\cite{prost_book,chandrasekhar_book,stephen1974,care2005}. 

Liquid crystals are fundamentally different from simple liquids in that they posses broken-symmetry variables as additional hydrodynamic variables~\cite{chaikin_book,forster_book,forster1974}.
In nematic liquid crystals, the local mean orientation of the molecules, termed as the \emph{director}, is the broken-symmetry variable~\cite{forster_book}.
The director is intimately coupled to  other hydrodynamic variables, making the hydrodynamic fluctuations in nematic 
liquid crystals different from that in simple fluids~\cite{durand1969,stephen1974}. 
For instance, the velocity fluctuations, which are isotropic in simple fluids, are coupled to the director fluctuations and become anisotropic in the nematic phase.
The director fluctuations are also responsible for the high light-scattering or turbidity of liquid crystals compared to simple liquids~\cite{stephen1974}. 
The static and dynamic correlations in the director fluctuations are often used to measure the elastic and viscous coefficients in experiments~\cite{durand1969,borsali1998,giavazzi2014}
and computer simulations~\cite{cleaver1991,humpert2015prl}.

The nature of hydrodynamic fluctuations in nematic liquid crystals has received a revived interest recently~\cite{humpert2015prl,humpert2015mol,humpert2016,chakrabarty2006}. 
The director fluctuations in nematic liquid crystals have long been observed to be diffusive in the incompressible limit~\cite{orsay1969,stephen1974}.
Recent simulation studies using Gay-Berne model of liquid crystals,
however, reveal that the fluctuations could either be diffusive or propagative, depending on the relative magnitude of the elastic and  viscous coefficients~\cite{humpert2015prl,humpert2015mol}.
The nature of the decay of hydrodynamic fluctuations are also in general closely related to the long-time behavior of single-particle autocorrelation functions~\cite{masters1998}.
Recent simulation studies indicate the existence of long-time power-law tails in the single-particle orientational autocorrelation functions in the nematic phase~\cite{humpert2016}.
The existence of long-time tails in the autocorrelation functions has also previously been reported in the context of isotropic-nematic transition~\cite{chakrabarty2006}. 
However, the evidence for the long-time tails in these studies were limited, primarily due to strong finite-size effects.

In this paper, we use the Lebwohl-Lasher model~\cite{lebwohl1972} to elucidate some of these aspects of hydrodynamic correlations in nematic liquid crystals.
The Lebwohl-Lasher model is simple in that the translational motion of the particles are arrested, and therefore allows
to analyze the hydrodynamic fluctuations arising purely from orientational motion of the molecules, as well as to study larger systems.
 In addition, the potential energy has a particularly simple form, which has been taken advantage of in theoretical evaluation of some of the elastic and 
thermodynamic properties of the system~\cite{priest1972,cleaver1991,ilg2012}.
We derive the corresponding linearized hydrodynamic equations.
The solutions of these linearized equations are then used to calculate the correlations in hydrodynamic fluctuations as well as angular momentum and orientational autocorrelation functions of tagged rotors.
The theoretical results are then compared with the results of molecular dynamics simulations.

The paper is organized as follows: Constitutive relations for the hydrodynamic currents are derived in Secs.~\ref{theory_I}-\ref{theory_II}.
Linearized solutions of the continuity equations are obtained in Sec.~\ref{theory_III}. Hydrodynamic correlations
are defined and explicit expressions are given in Sec.~\ref{theory_IV}. Simulation details are given in Sec.~\ref{sim_details}.
 Simulation results for the correlations functions are compared with the theoretical predictions in Secs.~\ref{sim_corr_fourier}-\ref{sim_corr_real}.
Sec.~\ref{conclusions} includes discussion and conclusions. Some details of the calculations are given in the appendices.

\section{Theory}\label{theory}
\subsection{Dynamic equations}\label{theory_I}
The local conservation of the energy  and angular momentum are expressed by the continuity equations
\begin{align} \label{e_cont}
\frac{\partial \epsilon}{\partial t} &= - \nabla\cdot \mathbf j^\epsilon~,\\
\intertext{and}
\frac{\partial \mbf l}{\partial t} &= - \nabla \cdot \boldsymbol \sigma~,
\label{l_cont}
\end{align}
where $\epsilon=\epsilon(\mbf r,t)$ and $\mbf l = \mbf l(\mbf r,t)$ are the time-dependent energy and angular momentum density, and $\mathbf j^{\epsilon}$ and $\boldsymbol \sigma$ are the corresponding currents, and 
$(\nabla\cdot\boldsymbol \sigma)_i=\nabla_j \mbf \sigma_{ij}$
 (we use the Einstein summation convention unless stated otherwise).
The thermodynamic state of the system in the isotropic phase is completely described by $\epsilon$ and $\mbf l$.
In the nematic phase, the local director $\mbf n$ forms an additional independent collective variable along with $\epsilon$ and $\mbf l$. 
The time evolution of the director $\mbf n$, which is not a conserved quantity, is given by~\cite{chaikin_book,forster_book} 
\begin{equation}
\frac{\partial \mbf n}{\partial t}=\boldsymbol \omega \times \mbf n - \mbf X'~,
\label{n_time}
\end{equation}
where $\boldsymbol\omega$ is the angular velocity field, which is related to the angular momentum density by $l_i=I_{ij} \omega_j$,
where $\mbf I$ is the moment of inertia tensor density. The first term on the rhs of Eq.~(\ref{n_time}) accounts for rigid rotations
and the quantity $\mbf X'$ for dissipative effects~\cite{forster1971,martin1972}. Since the director is normalized, $|\mbf n|=1$, we have $\mbf X'\cdot\mbf n=0$. 

The collective variables $\epsilon$, $\mbf l$ and $\mbf n$ are inter-related by the differential entropy density $s$ as~\cite{chaikin_book,martin1972}
\begin{equation}
Tds=d\epsilon-\boldsymbol \omega \cdot d\mbf l - h_{ij}d(\nabla_j n_i)~,
\label{GibbsDuhem}
\end{equation}
where $T$ is the temperature.
The quantity $h_{ij}$ is the  variable conjugate to the deformation $\nabla_j n_i$, and is given by $\frac{\partial f_e}{\partial (\nabla_j n_i)}$, where $f_e$ 
is the Frank free energy density~\cite{prost_book,chaikin_book}. 
The various elastic constants appearing in the general Frank free energy are all identical for the Lebwohl-Lasher model~\cite{cleaver1991,priest1972}, and therefore the free energy density takes the one-constant form $f_e=\frac{K}{2}\nabla_jn_i\nabla_jn_i$~\cite{prost_book},
where $K$ is the elastic constant. Consequently, for the Lebwohl-Lasher model we get
\begin{equation}
h_{ij}=K\nabla_j n_i~.
\label{h_LL}
\end{equation}

\subsection{Constitutive relations}\label{theory_II}
In order to obtain the constitutive relations 
for the currents $\mbf j^\epsilon$, $\boldsymbol \sigma$, and $\mbf X'$ in Eqs.~(\ref{e_cont})-(\ref{n_time}),  non-negativity of the entropy production is invoked. 
Using Eqs.~(\ref{e_cont})-(\ref{n_time}) in Eq.~(\ref{GibbsDuhem}), the time derivative of the total entropy of the system, $S=\int_{\mathcal V} d^dr s(\mbf r,t)$, can be written as (see Appendix~\ref{appendixI})
\begin{equation}
\frac{dS}{dt}=-\int_{\mathcal V} d^dr\frac{1}{T}\left[\frac{\mbf q}{T}\cdot\nabla T+(\boldsymbol\sigma + \mbf h'):\nabla\boldsymbol\omega+\mbf X'\cdot(\nabla\cdot\mbf h)\right]~,
\label{Sdot}
\end{equation}
where $\mbf q$ is the heat current  
\begin{equation}
\mbf q=\mbf j^\epsilon - \boldsymbol\sigma\cdot\boldsymbol\omega - \mbf h\cdot \mbf X'~, 
\label{heat_current}
\end{equation}
and  
\begin{equation}
h'_{ij}=\epsilon_{ikl}n^0_kh_{lj}~,
\label{h_prime}
\end{equation}
where $\epsilon_{ijk}$ is the Levi-Civita symbol, and we denote 
$\mbf a:\mbf b=a_{ij}b_{ji}$.
In obtaining Eq.~(\ref{Sdot}), 
the approximation $\boldsymbol \omega \times \mbf n\approx \boldsymbol \omega \times \mbf n^0$, where $\mbf n^0$ is the mean director,  
has been made in Eq.~(\ref{n_time}). From Eq.~(\ref{Sdot}) we infer for the entropy production to be non-negative, that  the currents need to be of the form~\cite{chaikin_book}
\begin{align}
\label{const_rel1}
\mbf q&=-\boldsymbol\kappa\cdot\nabla T ~,\\
\label{const_rel2}
\boldsymbol\sigma + \mbf h' &= -\boldsymbol\Gamma\cdot\nabla\boldsymbol\omega~,\\
\label{const_rel3}
\mbf X'&=-\gamma\nabla\cdot\mbf h~,
\end{align}
where $\kappa_{ij}$, $\Gamma_{ijkl}$, and $\gamma$ are positive semi-definite dissipative coefficients.

In the following, we provide explicit expressions for the currents $\mbf j^{\epsilon}$, $\boldsymbol \sigma$, and $\mbf X'$ for small fluctuations around the mean values of the collective variables.

Using Eqs.~(\ref{heat_current}), (\ref{const_rel1}) and (\ref{GibbsDuhem}) we get,
\begin{equation}
\mbf j^{\epsilon}\simeq \boldsymbol \kappa \cdot \nabla T \sim \nabla \epsilon~.
\end{equation}
The dynamics of $\epsilon$ is therefore decoupled from the rest of the collective variables $\mbf l$ and $\mbf n$, and will not be discussed further.

The dissipative coefficients $\Gamma_{ijkl}$ associated with the constitutive relation for $\boldsymbol \sigma$
can be written as (see Appendix.~\ref{appendixII})
\begin{align}
\Gamma_{ijkl}&=\delta_{jk}\left[\Gamma_\parallel n_in_l + \Gamma_\perp (\delta_{il}-n_in_l)\right]~,
\label{Gamma}
\end{align}
where $\Gamma_{\parallel,\perp}$ are constants. 
Substituting for $\boldsymbol\Gamma$ in  Eq.~(\ref{const_rel2}), and using Eq.~(\ref{h_prime}), we get
\begin{align}
\sigma_{ij}&=-h'_{ij}-\Gamma_{ijkl}\nabla_k\omega_l \nonumber \\
                   &\simeq -\epsilon_{ikl}n^0_kh_{lj}- \nabla_j\left[  \Gamma_\parallel \omega^{\parallel}_i + \Gamma_\perp \omega^{\perp}_i    \right]~,
		   \label{const_rel_sigma}
\end{align}         
where $\boldsymbol \omega^{\parallel}=(\mbf n\cdot\boldsymbol \omega)\mbf n$ 
and $\boldsymbol \omega^{\perp}=\boldsymbol \omega-(\mbf n\cdot\boldsymbol \omega)\mbf n$. We have also neglected second order terms of the form $\omega_l\nabla_jn_i$.

Finally, using Eq.~(\ref{const_rel3}) and (\ref{h_LL}) we get, 
\begin{equation}
\mbf X'=-\gamma K \nabla^2 \mbf n~.
\label{const_rel_xd}
\end{equation}
The viscous coefficient $\gamma$ is often termed rotational viscosity while $\Gamma_{\parallel,\perp}$ are known as spin viscosities.
\subsection{Linearized equations}\label{theory_III}
\subsubsection{Nematic phase}
We now consider the dynamics of $\mbf l$ and $\mbf n$ far from the isotropic-nematic transition point.
In the nematic phase it is convenient to decompose $\mbf l$ into components parallel and perpendicular
to the director: $\mbf l=\mbf l^{\parallel}+\mbf l^{\perp}$, with $\mbf l^{\parallel}=(\mbf n\cdot \mbf l)\mbf n$ and $\mbf l^\perp=\mbf l - (\mbf l\cdot \mbf n)\mbf n$.
Assuming uniaxial symmetry, the moment of inertia density will have the form~\cite{stark2005}, $I_{ij}=I_{\parallel}n_in_j+I_{\perp}\left[\delta_{ij}-n_in_j\right]$.
This implies, $\mbf l^\parallel=I^\parallel \boldsymbol\omega^\parallel$ and $\mbf l^\perp=I^\perp \boldsymbol\omega^\perp$.
Using these relations and inserting the constitutive relations Eqs.~(\ref{const_rel_sigma}) and (\ref{const_rel_xd}) in the dynamic equations Eqs.~(\ref{l_cont}) and (\ref{n_time}), we get 
\begin{align}
&\frac{\partial \mbf l^{\perp}}{\partial t} = K \mbf n^0 \times \nabla^2 \mbf n + \nu_{\perp} \nabla^2 \mbf l^{\perp}~,
\label{l_perp}\\
&\frac{\partial \mbf l^{\parallel}}{\partial t} = \nu_{\parallel} \nabla^2 \mbf l^{\parallel}~,
\label{l_para}\\
&\frac{\partial \mbf n}{\partial t}=I^{-1}_{\perp} \mbf l^\perp \times \mbf n^0 + \gamma K \nabla^2 \mbf n~,
\label{n_time_fin}
\end{align}
where $\nu_\perp=\Gamma_\perp/I_\perp$ and $\nu_\parallel=\Gamma_\parallel/I_\parallel$.
We note that for small fluctuations, Eqs.~(\ref{l_para})-(\ref{n_time_fin})  are identical to those derived in Refs.~\cite{lubensky2005,stark2005}
using the Poisson bracket formalism.

Writing $\mbf n=\mbf n^0+\delta \mbf n$ as before, and choosing $\mbf n^0$ along the $3$-direction in the Cartesian system ($\hat{\mbf e}_1,\hat{\mbf e}_2,\hat{\mbf e}_3$), for small fluctuations we have 
$\delta \mbf n \approx (n_1,n_2,0)$. This follows from the condition $\mbf n\cdot\delta \mbf n=0$ ($|\mbf n|=1$). 
Similarly, we have, $l^{\parallel}\approx l_3$ and $\mbf l^{\perp}\approx(l_1,l_2,0)$.
From Eqs.~(\ref{l_perp})-(\ref{n_time_fin}), we then find that the longitudinal component, $l_3$, is diffusive and is decoupled from the rest of the 
collective variables. The transverse components are coupled to the director fluctuations. 
Explicitly, $l_1$ is coupled to $n_2$, and $l_2$ to $n_1$. 
The coupled equations are readily solved using double Fourier-Laplace transform
\begin{equation}
\tilde {\mbf f}(\mbf k,s)=\int_{\mathcal V} d^dr \int_{0}^{\infty}\!dt\, \mbf f(\mbf r,t) e^{-i\mbf k\cdot \mbf r}e^{-st}~.
\end{equation}
Applying the transform for the ($l_x,n_y$) pair we get 
\begin{equation}
\left(
\begin{array}{c}
\tilde l_1(\mbf k,s)\\
\tilde n_2(\mbf k,s)
\end{array}
\right)=
\frac{1}{\Delta}
\left(
\begin{array}{cc}
s+\gamma K k^2 & Kk^2\\
-I^{-1}_\perp & s+\nu_\perp k^2
\end{array}
\right)
\left(
\begin{array}{c}
l_1(\mbf k,0)\\
n_2(\mbf k,0)
\end{array}
\right)~,
\label{matrix_eq}
\end{equation}
where $\Delta=\left(s-s_1\right)\left(s-s_2\right)$, with 
\begin{equation}
s_{1,2}=-\frac{1}{2}\nu_s k^2 \pm i\omega_s,
\label{poles}
\end{equation}
where the damping coefficient $\nu_s$ and frequency $\omega_s$ are given by
\begin{equation}
\nu_s=\gamma K + \nu_\perp
\label{nu_def}
\end{equation}
and
\begin{equation}
\omega_s = \left(KI^{-1}_{\perp}k^2 -\left(\gamma K -\nu_\perp\right)^2k^4/4 \right)^{1/2}.
\label{omega_s}
\end{equation}
Note that $\mbf f(\mbf k,0)$ denote variables which are only spatially Fourier transformed at $t=0$.
The corresponding equations for the pair $(l_2,n_1)$  are obtained by replacing $(l_1,n_2)\rightarrow(l_2,n_1)$ in Eq.~(\ref{matrix_eq}). 

Some comments on the frequency $\omega_s$ as given in Eqs.~(\ref{poles}) and (\ref{omega_s}) are in order.
The quantity $\omega_s$ could either be real or imaginary, depending on the relative magnitudes of the elastic constant and viscous coefficients, 
corresponding to diffusive or propagative transverse modes. 
In the limit, $(KI^{-1}_{\perp})^{1/2} \gg \left(\gamma K -\nu_\perp\right)k^2/2$, the modes are propagative with the dispersive frequency $\omega_s=\pm k(KI_\perp^{-1})^{1/2}$,
and damping factor $\frac{1}{2}\nu_sk^2$ (see Eq.~(\ref{poles})). In the opposite limit, we have diffusive modes with damping factors $\nu_\perp k^2$ and $\gamma K k^2$.
This limit is often assumed for nematic liquid crystals under normal experimental conditions~\cite{orsay1969,stephen1974}.
In what follows, we assume that $\omega_s$ is real, corresponding to propagating modes, as it turns out to be the case in the Lebwohl-Lasher model for small wavevectors.

\subsubsection{Isotropic phase}
In the isotropic phase, the broken symmetry variable $\mbf n$ vanishes and the linearized dynamics of the system is given by Eqs.~(\ref{l_perp}) and (\ref{l_para}), with $\nu_\perp=\nu_\parallel=\nu$~,
\begin{equation}
\frac{\partial \mbf l}{\partial t}=\nu \nabla^2 \mbf l~.
\end{equation}
Or, in the Fourier representation we have
\begin{equation}
\mbf l(\mbf k,t)=\mbf l(\mbf k,0)e^{-\nu k^2 t}.
\label{lk_soln}
\end{equation}

\subsection{Correlation functions}\label{theory_IV}
The autocorrelation functions of angular momentum and director fluctuations are defined as
\begin{equation}
C^{l}_{m}(\mbf k,t)=\langle l_m(\mbf k,t)l_m(-\mbf k,0)\rangle~,
\label{l_corr_def}
\end{equation}
and 
\begin{equation}
C^{n}_m(\mbf k,t)=\langle n_m(\mbf k,t)n_m(-\mbf k,0)\rangle~,
\label{n_corr_def}
\end{equation}
where $\langle \cdot \rangle$ denote the canonical ensemble average.
Note that the summation convention is not used in the definition of correlation functions.
In the isotropic phase, $C^l(\mbf k,t)$ can be readily obtained using Eq.~(\ref{lk_soln}), and is given by 
\begin{equation}
C^{l}_{m}(\mbf k,t)=\langle |l_m(\mbf k,0)|^2\rangle e^{-\nu k^2 t}~.
\label{corr_l_iso}
\end{equation}
In the nematic phase, $C^l_{1}$ can be obtained using Eq.~(\ref{matrix_eq}) by performing inverse Laplace transform:    
\begin{align}
C^{l}_{1}(\mbf k,t)&=\langle |l_1(\mbf k,0)|^2\rangle \mathcal L ^{-1} \left [\frac{s+\gamma K k^2}{(s-s_1)(s-s_2)}\right]\nonumber\\
&=\langle |l_1(\mbf k,0)|^2\rangle e^{-\frac{1}{2}\nu_s k^2 t}\nonumber\\
&~~~\times\left[\cos(\omega_s t)+\frac{(\gamma K-\nu_\perp) k^2}{2\omega_s}\sin(\omega_s t)\right]~,
\label{corr_l}
\end{align}
where the equilibrium relation $\langle l_1(\mbf k,0)n_2(-\mbf k,0)\rangle=0$ has been used. Similarly, the correlation function 
for the director fluctuations is given by  
\begin{align}
C^{n}_{2}(\mbf k,t)&=\langle |n_2(\mbf k)|^2\rangle e^{-\frac{1}{2}\nu_s k^2 t}\nonumber\\
&~~~\times\left[\cos(\omega_s t)-\frac{(\gamma K-\nu_\perp) k^2}{2\omega_s}\sin(\omega_s t)\right]~.
\label{corr_n}
\end{align}
Due to uniaxial symmetry around the nematic director, and the frequency $\omega_s$ and damping coefficient $\nu_s$ being independent of the direction of the wavevector $\mbf k$,
the correlation function $C^{n}_1$ is identical to $C^{n}_2$ and $C^{l}_2$ to $C^{l}_1$.

The equal-time correlations  $\langle |n_m(\mbf k)|^2\rangle$ and $\langle |l_m(\mbf k)|^2\rangle$ appearing in Eqs.~(\ref{corr_l_iso})-(\ref{corr_n}) can be obtained
using the equipartition theorem -- the free energy of the system is given by
\begin{align}
F&=\frac{1}{2}\int_{\mathcal V}d^dr \left [ K \nabla_j n_i\nabla_j n_i + \frac{l_\parallel ^2}{I_\parallel} + \frac{l_\perp ^2}{I_\perp} \right]\\
&=\frac{1}{2V}\sum_{\mbf k}\left[Kk^2|\mbf n(\mbf k)|^2 +\frac{|l_\parallel(\mbf k)| ^2}{I_\parallel} + \frac{|l_\perp(\mbf k)| ^2}{I_\perp} \right]. 
\end{align}
By equipartition of energy, we  get 
\begin{equation}
\begin{aligned}
&\langle |n_1(\mbf k)|^2\rangle=\langle |n_2(\mbf k)|^2\rangle=Vk_BT/(Kk^2)\\
&\langle |l_1(\mbf k)|^2\rangle=\langle |l_2(\mbf k)|^2\rangle=I_\perp V k_BT\\
&\langle |l_3(\mbf k)|^2\rangle=I_\parallel V k_BT\\
\end{aligned}
\label{equipartition}
\end{equation}
We assume that the principal moment of inertia of the rotors about the body-axis is zero and the components about axes perpendicular to the body-axis are equal. 
Then, in a perfectly ordered state, we have $I_\parallel=0$ and $I_\perp=\rho I$, where $\rho$ and $I $ are the number density and principal moment of inertia of the rotors.
On the other hand, in the isotropic state we have $I_\parallel=I_\perp=2I\rho /3$. This relation follows from that $I_{ij}$ is proportional to unit matrix in the disordered state, with 
the trace being $2I\rho$. We set the number density $\rho$ to unity. 

\section{Lebwohl-Lasher model}\label{LL_model}
The Hamiltonian of the Lebwohl-Lasher model is given by~\cite{lebwohl1972}
\begin{equation}
U=-\frac{\epsilon_0}{2} \sum_{\langle ij\rangle}^{N} P_2(\mbf u_i \cdot \mbf u_j)~,
\label{LL_Hamiltonian}
\end{equation}
where $\mbf u_i$ denote the unit orientation  vector of the $i$-th rotor, $P_2(x)=(3x^2-1)/2$ the second Legendre polynomial, $\epsilon_0$ is the strength of the interaction, $N$ is the number of rotors, 
and the summation extends over nearest-neighbor pairs.
The nematic-isotropic transition in the Lebwohl-Lasher model occurs at $k_BT_c/\epsilon_0\approx 0.56$ and $\approx 1.13$
for two- and three-dimensional cubic lattices~\cite{mondal2003,shekhar2012}.

\subsection{Simulation details}\label{sim_details}
We perform molecular dynamics simulations of a system of rotors on $L^d$ cubic lattices interacting via the Lebwohl-Lasher potential.
As previously stated, we consider linear rotors, i.e, the moment of inertia in the body-fixed frame of reference of a rotor 
has the form $\mbf I=\text{diag}(I,I,0)$, where $I$ is the principal moment of inertia. 
The  orientation $\mbf u_i$ and angular momentum $\mbf l_i$ of rotor $i$ evolve as 
\begin{equation}
\begin{aligned}
\dot{\mbf u}_i&=\boldsymbol \omega_i \times \mbf u_i~,\\
\dot{\mbf l}_i&=\boldsymbol \tau_i~, 
\label{eqn_motion}
\end{aligned}
\end{equation}
where $\boldsymbol \omega_i=\mbf I^{-1}\cdot\mbf l_i$ is the angular velocity and $\boldsymbol \tau_i$ the torque. The torque is given by~\cite{allen2006}
\begin{align}
\boldsymbol\tau_i&=-\sum_{j\neq i}\frac{\partial U}{\partial \left(\mbf u_i\cdot\mbf u_j\right)} \left(\mbf u_i \times \mbf u_j\right) \nonumber \\
               &=3\epsilon_0\sum_{j\neq i}\left(\mbf u_i\cdot\mbf u_j\right) \left(\mbf u_i \times \mbf u_j\right)~,
\end{align}
where we used Eq.~(\ref{LL_Hamiltonian}) in the second step.

The equations of motion Eqs.~(\ref{eqn_motion}) are integrated numerically using a leap-frog algorithm~\cite{allen_book}.
The initial orientations of the rotors in the nematic phase are generated using Monte-Carlo simulations, and an equal rotation 
is applied to each rotor such that the initial director points along the $\hat{\mbf e}_3$ axis. The length of the simulation runs is chosen such that the drift in the director from the initial direction is negligible.
The initial angular velocities are Maxwell-Boltzmann distributed. Note that each rotor has only two rotational degrees of freedom and the component of 
angular velocity parallel to the rotor vanishes. In order to avoid a rigid rotation of the whole system and therefore of the director, the total angular momentum is initialized to zero.
A simple rescaling of the angular velocities is applied at every time step in order to maintain the system at a constant temperature.
We specify the temperature, time and length scales in reduced units $T^*=k_BT/\epsilon_0$, $t^*=t/(I/\epsilon_0)^{1/2}$, $r^*=r/a$, where $a$ is the lattice spacing.
All the measured quantities are reported in these units. 

In the following, we define the microscopic expressions for the hydrodynamic fields in the simulations. 
The nematic director fluctuations are conveniently described in terms of the local order parameter tensor~\cite{forster1974,forster_book}
\begin{equation}
Q_{lm}(\mbf r,t)=\frac{1}{N}\sum_{i=1}^{N}\left(u_{il}(t)u_{im}(t)-\frac{1}{3}\delta_{lm}\right)\delta(\mbf r-\mbf r_i)~,
\end{equation}
where $\mbf r_i$'s are position coordinates of the rotors and $l,m=1,2,3$. 
The macroscopic director $\mbf n^0$ is parallel to the eigenvector of $\int_{\mathcal V} d^dr Q_{lm}$ corresponding to the largest eigenvalue, 
with the eigenvalue being related to the order parameter $S$ as $2S/3$.
For $\mbf n^0$ pointing along the $\hat{\mbf e}_3$ axis, the local director components are given in term of $Q_{lm}$ as $n_1=Q_{13}/S$ and $n_2=Q_{23}/S$~\cite{forster1974,forster_book}.
The order parameter tensor in the Fourier representation reads 
\begin{equation}
Q_{lm}(\mbf k,t)=\frac{1}{N}\sum_{i=1}^{N}\left(u_{il}(t)u_{im}(t)-\frac{1}{3}\delta_{lm}\right)e^{-i\mbf k\cdot \mbf r_i}~,
\end{equation}
where the wavevector $\mbf k=(2\pi/L)(\kappa_1,\kappa_2,\kappa_3)$, where $\kappa_1,\kappa_2,\kappa_3$ are integers.
Similarly, the angular momentum field is defined as 
\begin{equation}
\mbf l(\mbf k,t)=\sum_{i=1}^N \mbf l_i(t) e^{-i\mbf k\cdot\mbf r_i}~.
\label{lk_def}
\end{equation}
The corresponding longitudinal and transverse components are then given by $(0,0,l_3(\mbf k,t))$ and $(l_1(\mbf k,t),l_2(\mbf k,t),0)$, respectively.

\subsection{Correlations in k-space}\label{sim_corr_fourier}

\begin{figure}
\includegraphics[width=\columnwidth]{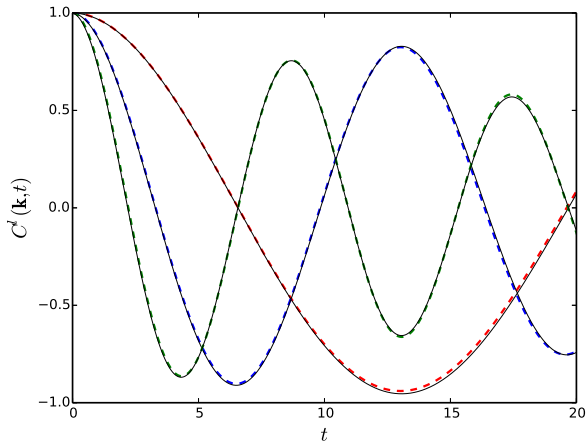}
\caption{Angular momentum autocorrelation functions (normalized) as defined in Eq.~(\ref{l_corr_def}) for components perpendicular to the nematic director for different wavevectors. 
The dashed lines corresponds to the simulation results and the solid lines are fits to the theoretical expression in Eq.~(\ref{corr_l}). 
Parameters are chosen as $T=1.0$, $N=32^3$, and $k=2\pi/32$, $4\pi/32$ and $6\pi/32$ (top to bottom at $t=2.5$).
}
\label{kspace_corr}
\end{figure}
We now calculate in the simulations the autocorrelation functions of the angular momentum and director fluctuations as defined in Eqs.~(\ref{l_corr_def}) and (\ref{n_corr_def}). 
In evaluating Eq.~(\ref{n_corr_def}), we use the definition $n_1(\mbf k,t)=Q_{13}(\mbf k,t)/S$ and $n_2(\mbf k,t)=Q_{23}(\mbf k,t)/S$, as previously stated. 
Figure~\ref{kspace_corr} displays the transverse angular momentum correlations in the nematic phase, for small wavevectors. 
In agreement with Eq.~(\ref{corr_l}), the fluctuations in the transverse components decay propagatively, with the 
frequency being determined by the Frank elastic constant and the damping by a combination of the elastic constant and viscous coefficients.
Similarly, the director fluctuations also decay as propagating modes  (not shown) in agreement with Eq.~(\ref{corr_n}).
The correlations are independent of the direction of the 
wavevector $\mbf k$, and therefore the decay of splay ($n_1,\hat {\mbf k}=\hat {\mbf e}_1$), twist ($n_2,\hat {\mbf k}=\hat {\mbf e}_1$) and bend ($n_1,n_2,\hat {\mbf k}=\hat {\mbf e}_3$) 
fluctuations show identical decay as expected.

\begin{table}[t]
\begin{ruledtabular}
\begin{tabular}{ccccccc}
 $T $ & $S$ & $I_\perp $ & $K $  & $\nu_s $ \\
\hline
 0.10 & 0.975 & 0.992 & 2.808  &  -- \\
 0.30 & 0.921 & 0.974 & 2.587  &  -- \\
 0.40 & 0.892 & 0.964 & 2.520  &  -- \\
 0.75 & 0.767 & 0.922 & 1.988  & 0.022 \\
 1.00 & 0.604 & 0.863 & 1.293  &  0.194  \\
\end{tabular}
\end{ruledtabular}
\caption{\label{table1}
Temperature dependence of the order parameter ($S$), principal moment of inertia about axis perpendicular to the director ($I_\perp$), Frank elastic constant ($K$), and  
damping coefficient ($\nu_s$) defined in Eq.~(\ref{nu_def}). $I_\perp$ was obtained directly from simulations, whereas $K$ and $\nu_s$ were obtained by fitting Eq.~(\ref{corr_l}) against 
the correlation function obtained from the simulations. Typical error in the data presented is less than $1\%$, except for $\nu_s$, which is less than $3.5\%$. System size: $N=32^3$.
}
\end{table}

The Frank elastic constant $K$ and the viscous coefficient $\nu_s$ can be obtained by fitting the theoretical expressions Eq.~(\ref{corr_l}) or Eq.~(\ref{corr_n}) against the 
time correlation functions obtained from the simulations. The so-obtained values are listed in Table I. We find good agreement between our results for $K$ and that obtained in Ref~\cite{cleaver1991} using static 
director fluctuations. To our knowledge, the viscous coefficient $\nu_s=\gamma K + \nu_\perp$ for the Lebwohl-Lasher model has not yet been evaluated. 
We find that $\nu_s$ is smaller than $K$ by orders of magnitude, and decreases as the temperature is lowered from the isotropic-nematic transition point. 
We note that the relative smallness of spin and rotational viscosities has previously been observed in fluids composed of linear molecules~\cite{evans1978,hansen2009}. 

\begin{figure}
\includegraphics[width=\columnwidth]{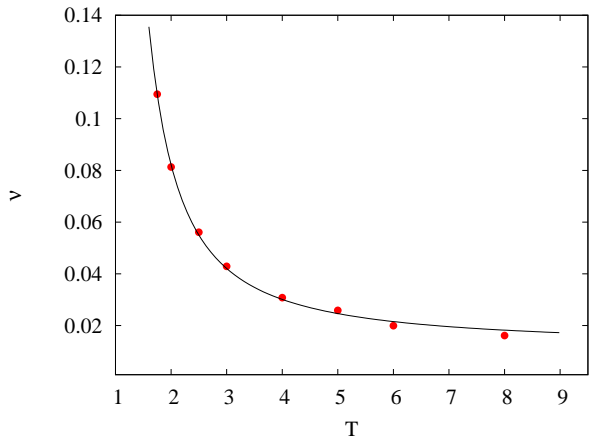}
\caption{Temperature dependence of spin viscosity $\nu$ in the isotropic phase.
Symbols denote the simulation results and the line corresponds to the fitting function given by Eq.~(\ref{nu_fit}).
System size: $N=32^3$.
}
\label{viscosity}
\end{figure}

In the isotropic phase, the angular momentum fluctuations decay diffusively, in agreement with Eq.~(\ref{corr_l_iso}).
Figure \ref{viscosity} displays the temperature dependence of the spin viscosity obtained by fitting Eq.~(\ref{corr_l_iso}) against the simulation results for the correlation functions.
The viscosity obtained is well described by the Arrhenius relation~\cite{imura1972}
\begin{equation}
\nu=\nu_0 \exp(E/k_BT))~,
\label{nu_fit}
\end{equation}
where $\nu_0$ and $E$ are constants. We find $\nu_0\approx 0.01$ and $E\approx 4.02$.
The fitted value of the activation energy $E$ is consistent with the fact that the nearest-neighbor interactions contribute, within orders of magnitude, $6 \epsilon_0$ to $E$. 
Similarly, $\nu(T_c)$ obtained using Eq.~(\ref{nu_fit}) is of the order of $\frac{1}{2}\nu_s$, close the transition, as expected. 
Arrhenius-like dependence of viscosity on temperature in the isotropic phase has previously been observed in experiments~\cite{imura1972}.

\subsection{Long-time tails}\label{sim_corr_real}
\begin{figure}
\includegraphics[width=\columnwidth]{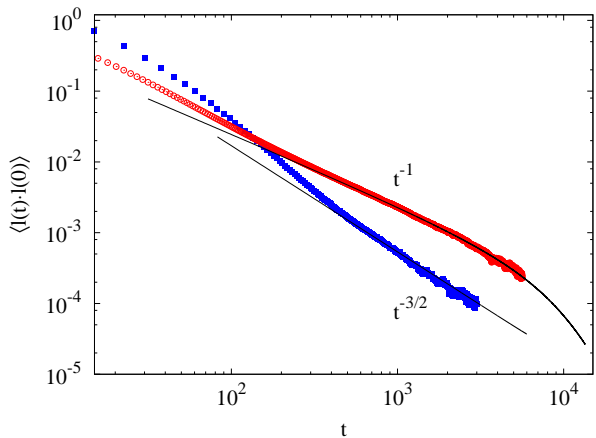}
\caption{Single-particle angular momentum autocorrelation functions in the real space in the isotropic phase. Simulation results are represented by symbols (circles - $2d$, squares - $3d$) and theoretical 
results (Eq.~\ref{sum}) by solid lines. For $2d$ systems, while the position coordinates of the rotors are confined to a plane, the orientation vectors point on a $3d$ unit sphere.
Parameters: $N=64^3$, $T=2.63$ for $3d$ and $N=100^2$, $T=1.50$ for $2d$ systems.   
}
\label{tail}
\end{figure}
We now consider the long-time behavior of single-particle autocorrelation functions (ACFs) in the real-space. 
The ACF of the angular momentum density is given by
\begin{align}
C^{l}(\mbf r,\mbf r', t)&=\langle \mbf l(\mbf r,t)\cdot \mbf l(\mbf r',0)\rangle\nonumber\\
                        &=\frac{1}{V^2}\sum_{\mbf k}\langle \mbf l(\mbf k,t)\cdot\mbf l(-\mbf k,0)\rangle e^{i\mbf k\cdot(\mbf r-\mbf r')}~.
		       \label{corr_l_sum}
\end{align}
Similarly, the director fluctuation ACF is  given by
\begin{align}
C^{n}(\mbf r,\mbf r', t)&=\langle \mbf n(\mbf r,t)\cdot \mbf n(\mbf r',0)\rangle\nonumber\\
                        &=\frac{1}{V^2}\sum_{\mbf k}\langle \mbf n(\mbf k,t)\cdot\mbf n(-\mbf k,0)\rangle e^{i\mbf k\cdot(\mbf r-\mbf r')}~.
		       \label{corr_n_sum}
\end{align}
Here, the summation is over all nonzero wavevectors.
Since the rotors are fixed in space, the ACFs of individual rotors can be directly obtained from the ACFs in the hydrodynamic fields.
Explicitly, the ACFs of rotor $i$, say, can be obtained by setting $\mbf r=\mbf r'=\mbf r_i$ in Eqs.~(\ref{corr_l_sum}) and (\ref{corr_n_sum}).
We denote $c^l(t)$ and $c^n(t)$ for the single-particle angular momentum and orientational ACFs. 

\subsubsection{Isotropic phase}

In the isotropic phase, $c^l(t)$ is obtained by using Eqs.~(\ref{l_corr_def}) and (\ref{corr_l_iso}) in Eq.~(\ref{corr_l_sum}), and is given by
\begin{align}
c^l_\text{iso}(t)&=\langle \mbf l_i(t)\cdot\mbf l_i(0)\rangle\nonumber\\
        &=\frac{2Ik_BT}{V} \sum_{\mbf k} e^{-\nu k^2 t}~,
	\label{sum}
\end{align}
where we used, $\langle |l(\mbf k,0)|^2\rangle=2IVk_BT$ (see Eq.~(\ref{equipartition})). For large system sizes, the summation can be approximated by integration, yielding
\begin{equation}
c^{l}_\text{iso}(t)\simeq\frac{2Ik_B T}{(4\pi \nu t)^{d/2}}~.
\label{lt_tail}
\end{equation}
We therefore find the long-time $t^{-d/2}$  hydrodynamic decay for infinite systems. Obviously, we 
expect deviations from $t^{-d/2}$ decay for finite systems.
Figure~\ref{tail} shows a comparison of the theoretical and simulation results for 
the single-particle angular momentum ACF  for two- and three-dimensional lattices.
The theoretical results are obtained by numerical summation of Eq.~(\ref{sum}), with the viscosity $\nu$ obtained from the k-space correlation function (Eqs.~(\ref{l_corr_def}) and (\ref{corr_l_iso}))
for the smallest wavevector, $k_0=2\pi/L$. In agreement with theory, the autocorrelation function decay as $t^{-d/2}$ over a long-time window. The decay becomes exponential for longer times
$t>(\nu k_0^2)^{-1}$. For such times, the summation in Eq.~(\ref{sum}) is dominated by the term $e^{-\nu k_0^2t}$, leading to an exponential decay, a finite size effect.
For short times, all the wavevectors $k<1/(\nu t)^{1/2}$ contribute to the summation. However, the $e^{-\nu k^2t}$ form of the summand is valid only for small wavevectors, leading to deviation of 
simulation results from the theoretical prediction. The short-time behavior may be described by treating the viscosity $\nu$ wavevector dependent for large wavevectors, however, this goes beyond the scope of this manuscript.

\subsubsection{Nematic phase}
In the nematic phase, the orientational ACF is obtained using Eqs.~(\ref{n_corr_def}) and (\ref{corr_n}) in Eq.~(\ref{corr_n_sum}). The long-time behavior of the 
ACF is given by
\begin{align}
c^{n}_\text{nem}(t)&=\langle \mbf u^{\perp}_i(t)\cdot\mbf u^{\perp}_i(0)\rangle\nonumber\\
        &\simeq\frac{2k_BT}{KV}\sum_{\mbf k}\frac{1}{k^2} e^{-\frac{1}{2}\nu_s k^2 t}\cos(k c t)~,
\end{align}
where $c=(K/I_\perp)^{1/2}$. Approximating the summation by integration as before, for  three-dimensional systems we get
\begin{equation}
c^{n}_\text{nem}(t)\simeq\frac{k_BT}{K\sqrt{2\pi^3\nu_s t}} e^{-\frac{c^2}{2\nu_s}t}~.
\label{c_n_nem}
\end{equation}
A similar calculation for the  angular momentum ACF in the nematic phase yields
\begin{align}
c^{l}_\text{nem}(t)\simeq \frac{k_BTI_\perp}{\sqrt{2\pi^3}} & \left[\left(\nu_s t\right)^{-\frac{3}{2}}-\left(\frac{c}{\nu_s}\right)^2\left(\nu_s t\right)^{-\frac{1}{2}}\right]\nonumber \\
&\times e^{-\frac{c^2}{2\nu_s}t}~.
\label{c_l_nem}
\end{align}
The long-time tails in the orientational and angular momentum ACFs in the nematic phase are therefore exponentially suppressed.
Similar strong decay of ACFs in the Lebwohl-Lasher model, close to the isotropic-nematic transition~\cite{chakrabarty2006} and in the Gay-Berne model in the nematic phase~\cite{humpert2016} has been previously reported.
Since $c^2/\nu_s = K/(I_\perp \nu_s) \gg 1$ for the Lebwohl-Lasher model, the ACFs have vanishing values for times where Eqs.~(\ref{c_n_nem}) and (\ref{c_l_nem}) are valid ($t\sim (\frac{1}{2}\nu_s k_0^2)^{-1}$).
It is therefore difficult to compare the theoretical expressions Eqs.~(\ref{c_n_nem}) and (\ref{c_l_nem}) with the simulations results. However, in the simulations we observe that 
the ACFs in the nematic phase decay rapidly so that no power-law regime could be identified.

\section{Conclusion}\label{conclusions}

The collective modes in the Lebwohl-Lasher model, as in the general case of rotors on lattices, consist of the director fluctuations (two transverse),
angular momentum fluctuations (two transverse, one longitudinal), and the energy fluctuations (one).
In the nematic phase, the director fluctuations are coupled to the transverse angular momentum fluctuations, leading to similar temporal decay. 
The fluctuations decay propagatively,  with the frequency in general being non-dispersive. For small wavevectors, the frequency becomes dispersive and 
proportional to the square root of the Frank elastic constant. The damping coefficient depends on the Frank elastic constant, spin viscosity, and rotational viscosity.
Moreover, the frequency and damping coefficient are independent of the direction of the wavevector. It is a consequence of the invariance of the Hamiltonian of the Lebwohl-Lasher model under simultaneous rotation of the  
rotors. Our results complement the recent finding~\cite{humpert2015prl,humpert2015mol} that propagating modes do exist in nematic liquid crystals in certain parameter range,
against the long-standing experience of observing only diffusive modes.

In the isotropic phase, however, the fluctuations are diffusive. The diffusive nature of the fluctuations interestingly manifest itself as long-time power-law tails in the 
single-particle autocorrelation functions. In particular, the angular momentum autocorrelation functions decay as $t^{-d/2}$ for long-times for infinite systems with $d$ spatial dimensions.
Note that this is in contrast with the well known $t^{-(d/2+1)}$ decay of angular momentum autocorrelations of particles immersed in simple fluids~\cite{hauge1973,masters1985,lowe1995}.
The reason for the difference is apparent -- the dynamic equations of the angular momentum density 
in flowing nematic fluid are different from that in a lattice system where the linear motion of the particles are arrested. 

The correlations in the director fluctuations in the nematic phase do not show long-time power-law behavior in three-dimensions.
Since the director fluctuations in the Fourier space are propagative, i.e., damped sinusoidally, 
the correlations in real space are suppressed by these oscillations. The anticipated power-law $t^{-1/2}$~\cite{masters1998}, which is based on the assumption that the modes are diffusive,
is exponentially suppressed even for infinite systems. 
 
The correlation functions derived here provide an alternate route to measure the Frank elastic constant and viscous coefficients. 
The simulation results for the Frank elastic constant obtained from the correlation functions are in good agreement with previous results obtained
using different methods. The viscous coefficients are also evaluated. In particular, the spin viscosity in the isotropic phase
follows Arrhenius-like dependence on temperature. The spin and rotational viscosity in the nematic phase are smaller than the Frank elastic constant by orders of magnitude. 
The study presented here can be directly used or adapted to study hydrodynamic correlations in lattice rotors with general interaction potentials.

\begin{acknowledgments}
We acknowledge support from the 7th framework program of the European Union via MC--CIG Grant No.~631233. 
\end{acknowledgments}

\appendix
\section{Entropy production}\label{appendixI}
Using Eqs. (\ref{e_cont})-(\ref{n_time}) in Eq.~(\ref{GibbsDuhem}), the time evolution of the entropy density is given by
\begin{equation}
T\frac{\partial s}{\partial t}=-\nabla\cdot \mbf j^\epsilon + \boldsymbol \omega\cdot(\nabla\cdot\boldsymbol\sigma) + h_{ij}\nabla_{j}(X'_i-\epsilon_{ikl}\omega_kn_l)~.
\label{entropy1}
\end{equation}
Some rearrangements of the terms in the above equation are in order: 
\begin{eqnarray*}
\boldsymbol \omega\cdot(\nabla\cdot\boldsymbol\sigma)&=&\omega_i\nabla_j\sigma_{ij}\\
&=&\nabla_j(\omega_i\sigma_{ij})-\sigma_{ij}\nabla_{j}\omega_i\\
&=&\nabla\cdot(\boldsymbol\sigma\cdot\boldsymbol\omega)-\boldsymbol\sigma:\nabla\boldsymbol\omega
\end{eqnarray*}
Similarly, 
\begin{eqnarray*}
h_{ij}\nabla_jX'_i&=&\nabla_j(h_{ij}X'_i)-X'_i\nabla_j h_{ij}\\
&=&\nabla\cdot(\mbf h\cdot\mbf X')-\mbf X'\cdot(\nabla\cdot \mbf h)
\end{eqnarray*}
And, 
\begin{eqnarray*}
h_{ij}\nabla_{j}(\epsilon_{ikl}\omega_kn_l)&\approx& \epsilon_{ikl}h_{ij}n^0_l\nabla_{j}\omega_k\\
&=& \mbf h':\nabla \boldsymbol \omega
\end{eqnarray*}
Here, we assumed small $\boldsymbol\omega$ limit and neglected the second order term by writing $\mbf n=\mbf n^0+\delta \mbf n$, where $\mbf n^0$ is the macroscopic director, and defined $h'_{ij}=\epsilon_{ikl}n^0_kh_{lj}$.
Using these simplifications Eq.(\ref{entropy1}) can be written as
\begin{equation}
T\frac{\partial s}{\partial t}=-\nabla\cdot \mbf q-(\boldsymbol\sigma + \mbf h'):\nabla\boldsymbol\omega-\mbf X'\cdot(\nabla\cdot\mbf h)
\label{S_dot}
\end{equation}
where the heat current $\mbf q=\mbf j^\epsilon - \boldsymbol\sigma\cdot\boldsymbol\omega - \mbf h\cdot\mbf X'$.
Dividing Eq.~(\ref{S_dot}) by $T$ and integrating over the volume and assuming that $\mbf q$ vanishes at the boundary, we get
\begin{equation}
\frac{dS}{dt}=-\int d^dr\frac{1}{T}\left[\frac{\mbf q}{T}\cdot\nabla T+(\boldsymbol\sigma + \mbf h'):\nabla\boldsymbol\omega+\mbf X'\cdot(\nabla\cdot\mbf h)\right]~,
\end{equation}
where we have used integration by parts for the first term in the rhs.

\section{Friction coefficients $\Gamma_{ijkl}$}{\label{appendixII}}
The most general form of $\boldsymbol\Gamma$ with uniaxial symmetry is
\begin{align}
\Gamma_{ijkl}&=a_1 \delta_{ij}\delta_{kl} + a_2 \delta_{ik}\delta_{jl} + a_3 \delta_{il}\delta_{jk} \nonumber \\ 
             &+a_4 \delta_{ij}n_kn_l+a_5 \delta_{ik}n_jn_l+a_6 \delta_{il}n_jn_k \nonumber \\
	     &+a_7 \delta_{jk}n_in_l+a_8 \delta_{jl}n_in_k+a_9 \delta_{kl}n_in_j \nonumber \\
	     &+a_{10} n_in_jn_kn_l~,
	     \label{Gamma_gen}
\end{align}
where $a_i$'s are constants, $\mbf n$ is the nematic director. The odd terms in $\mbf n$ are absent due to $\mbf n\rightarrow -\mbf n$ invariance of nematic phase.
From Eqs.~(\ref{Sdot}) and (\ref{const_rel2}), the contribution to entropy production density due to terms involving $\boldsymbol\Gamma$ is 
\begin{align}
\dot s_\Gamma &= (\boldsymbol \sigma + \mbf h'):\nabla \boldsymbol \omega \\
       &= \Gamma_{ijkl}\nabla_j \omega_i \nabla_k\omega_l
\label{s_dot}
\end{align}
Now, consider the transformation where all the rotors are rotated by same angle, without rotating the lattice vectors, i.e., 
\begin{align}
            \mbf n' &= \mathcal R \mbf n ~~,~~\boldsymbol \omega' = \mathcal R \boldsymbol \omega~,\\
	    \boldsymbol \Gamma' &=\boldsymbol \Gamma(\mbf n')~~,~~\nabla' =\nabla~,
\end{align}
where $\mathcal R$ is a rotation matrix. Since the entropy production should be invariant under overall rotations, from Eq.~(\ref{s_dot}) we  get
\begin{equation}
\Gamma_{ijkl}'\nabla_j \omega_i' \nabla_k\omega_l'=\Gamma_{ijkl}\nabla_j \omega_i \nabla_k\omega_l
\end{equation}
Or,
\begin{equation}
\Gamma_{ijkl}'\mathcal R_{im}\mathcal R_{ln}\nabla_j \omega_m \nabla_k \omega_n=\Gamma_{ijkl}\nabla_j \omega_i \nabla_k\omega_l
\end{equation}
This amounts to say 
\begin{equation}
\Gamma_{ijkl}'\mathcal R_{im}\mathcal R_{ln}=\Gamma_{mjkn}
\end{equation}
This implies, since the indices $j$ and $k$ are intact in the above equation, coefficients of terms with $n_j'$ and/or $n_k'$ in Eq.~(\ref{Gamma}) will vanish.
The coefficients $a_1$ and $a_2$ should also vanish, leaving only $a_3$ and $a_7$ non-vanishing. Defining $a_3=\Gamma_\perp$ and $a_3+a_7=\Gamma_\parallel$, we get
\begin{align}
\Gamma_{ijkl}&=a_3\delta_{il}\delta_{jk} + a_7 \delta_{jk}n_in_l\\
             &=\delta_{jk}\left[\Gamma_\parallel n_in_l + \Gamma_\perp (\delta_{il}-n_in_l)\right]
\end{align}

\end{document}